# The Pierre Auger Observatory
# Progress and First Results

Paul Mantsch for the Pierre Auger Collaboration[a,b].
*(a) Observatorio Pierre Auger, Av. San Martin Norte 304, 5613 Malargűe, Argentina*
*(b) Pierre Auger Collaboration: http://www.auger.org/admin/ICRC_2005_Author_List.pdf*
Presenter: Paul Mantsch (mantsch@fnal.gov)

The Pierre Auger Observatory was designed for a high statistics, full sky study of cosmic rays at the highest energies. Energy, direction and composition measurements are intended to illuminate the mysteries of the most energetic particles in nature. The Auger Observatory utilizes a surface array together with air fluorescence telescopes which together provide a powerful instrument for air shower reconstruction. The southern part of the Auger Observatory, now under construction in the Province of Mendoza, Argentina, is well over half finished. Active detectors have been recording events for one and a half years. Preliminary results based on this first data set are presented.

## 1. Introduction

Cosmic rays with energies near $10^{20}$ eV have been a continuing mystery since John Linsley reported the first such event in 1963[1]. As yet there are no known sources and no known mechanism for accelerating particles to these energies. Interaction with the cosmic microwave background (CMB) further constrains, protons, the most likely candidates, to come from distances not greater than about 50 Mpc.

The Pierre Auger Observatory was conceived in 1991 to probe the mystery of the highest energy cosmic rays with high statistics. Shortly afterwards, very high energy events were reported by the AGASA array (>2 x $10^{20}$ eV) [2] and the Fly's Eye (>3 x $10^{20}$ eV) [3] stimulating further interest. A collaboration consisting of a diverse group of 360 scientists and engineers from 60 institutions in 16 countries was assembled to construct the world's largest air shower detector designed to identify the origin of the highest energy cosmic rays and perhaps understand the acceleration mechanism. The Auger Observatory will occupy two sites, one in the northern hemisphere and one in the southern hemisphere. Construction at the first of the two sites is now underway in a remote location at the base of the Andes Mountains near Malargüe, Argentina.

## 2. Objectives

The first clue to the nature of the source would be a measurement of the spectrum with sufficient precision to determine if the flux above about $10^{19.8}$eV is suppressed by energy loss due to interactions with the CMB. This feature of the spectrum is known as the GZK cutoff [4, 5]. A continuing spectrum would suggest local or perhaps exotic origin. The first objective of the Observatory, then, is to measure in detail the features of the spectrum in the region of the GZK cutoff.

Arrival direction distributions with sufficient precision and statistics may reveal the presence of anisotropies and possibly point sources.

Another clue to the origin of cosmic rays is their composition. Features of the showers as measured by the Auger Observatory can distinguish between light and heavy nuclei on a statistical basis. It is also possible to use distinctive signatures to identify showers initiated by photons, neutrinos and perhaps even exotic particles. For example, because of their deep penetration in the atmosphere and muon deficiency, showers from photon primaries can be distinguished from those of nucleonic primaries. The identification of significant numbers of photon primaries is a smoking gun for most models that tie the highest energy cosmic ray to exotic origins such as decay of primordial relics.

In summary the primary design objectives for the Auger Observatory are to measure the energy, direction and composition with a large aperture air shower detector (>7000 km$^2$ sr per site above $10^{19}$ eV) in both hemispheres for high statistics and uniform, full sky exposure.

## 3. The Design

An extremely powerful feature of the Auger design is the capability of observing air showers simultaneously by two different but complementary techniques. On dark moonless nights, air fluorescence telescopes record the development of what is essentially the electromagnetic shower that results from the interaction of the primary particle with the upper atmosphere. The surface array measures the particle densities as the shower strikes the earth just beyond its maximum development. By recording the light produced by the developing air shower, fluorescence telescopes can make a near calorimetric measurement of the energy. This energy calibration can then be transferred to the surface array with its 100% duty factor and large event gathering power. As shown later, this energy conversion and subsequent determination of the spectrum can be done with minimal reliance on numerical simulations or on assumptions about the composition or interaction models.

Independent measurements by the surface array and the fluorescence detectors alone have limitations that can be overcome by comparing the results of their measurements. Indeed the hybrid measurements enforce a discipline in understanding the details of the capabilities and associated systematic uncertainties of the two techniques. The comparison of the results of the two techniques that lead to divergent results could also reveal interesting physics such as insight into hadronic interactions at these extreme energies.

When used together the surface array and the fluorescence detector can make a precision energy and direction measurement. The improved angular precision of hybrid events (~0.6º) makes the hybrid event especially valuable for anisotropy studies. Even when only one surface detector station triggers it can provide timing information to the fluorescence measurement that results in a hybrid shower geometrical reconstruction.

The two techniques also contribute a set of complementary mass-sensitive parameters. Both the depth and fluctuation of shower maximum are measures of composition. Heavy primaries tend to shower early, fluctuate less and produce more muons. Other composition driven features of shower development show up in the surface detectors as variations in the pulse rise time, radius of curvature and thickness of the shower front.

A combination of a self-imposed total cost limitation and the size of suitable sites led to an array of particle detectors occupying 3000 km$^2$. A detector spacing of 1.5 km was selected to ensure full efficiency for recording events above $10^{19}$ eV. (In reality we have been able to achieve full efficiency down to 3 x $10^{18}$ eV for zenith angles less than 60°). While at four locations around the perimeter of the surface array, 24 fixed optical telescopes view the sky over the array. On dark, cloudless nights the faint fluorescence light is captured from the developing air showers in what is effectively the transparent calorimeter formed by the atmosphere. The Observatory Plan is shown in Figure 1.

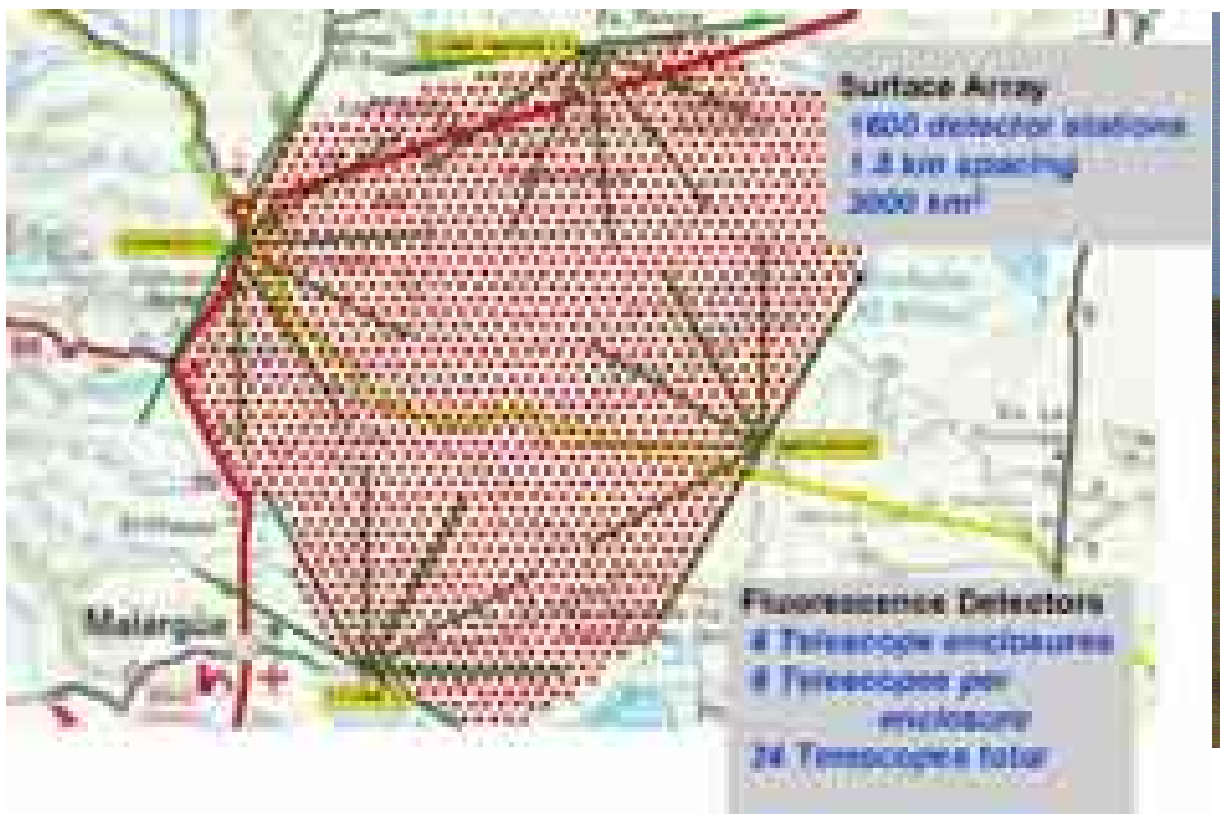

**Figure 1.** The Auger Observatory plan.

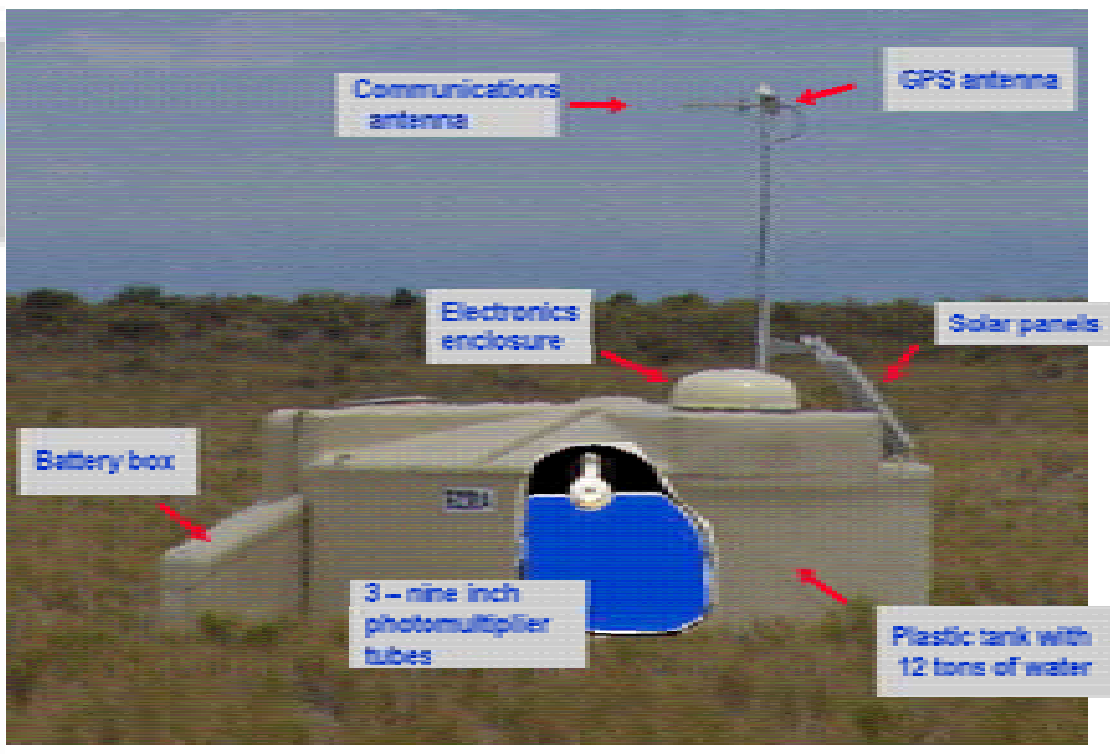

**Figure 2.** Details of the surface detector station.

The water Cherenkov particle detector was chosen for use in the surface array because of robustness and low cost. It is designed to withstand the harsh environment of the Pampa Amarilla in western Argentina. Sand, salt, rain, snow and hail are sometimes blown by winds up to 160 km/hour. Excursions of $20^{\circ}$ C in temperature in a day are not uncommon. The surface detector station consists of a 12,000 liter rotomolded high density polyethylene water tank containing a sealed laminated polyethylene liner with a reflective inner surface. The outer layer of the tank is colored light beige to blend with the surrounding desert. Cherenkov light from the passage of particles is collected by three nine inch photomultiplier tubes that look through windows of clear polyethylene into highly purified water. The surface detector station is self contained. A solar system provides 10 watts of power for the PMTs and electronics package. The electronics package, consisting of a processor, GPS receiver, radio transceiver and power controller, is mounted on the tank under an aluminum dome. The components of the surface detector station are shown in Figure 2.

The details of the fluorescence detector telescope are shown in Figure 3. The most important feature that differentiates the Auger design from that of earlier fluorescence detectors is the use of Schmidt optics. The 11 $m^2$ segmented spherical mirror focuses light from the aperture window on to a camera containing 440 photomultipliers. The aperture stop with a partial corrector lens controls the spherical aberration of the image. An optical filter, which passes nitrogen fluorescence light, also serves as a window over the aperture. Fully enclosed, the space containing the telescopes and electronics can be kept clean and climate controlled. Figure 4 contains two views of one of the fluorescence detector enclosures containing six telescopes, each subtending 30º vertically and horizontally. The shutters seen in the left-hand figure close automatically when the wind becomes too high or rain is detected. All functions of the fluorescence enclosures and telescopes are operated remotely from the data acquisition room on the Auger campus. The antennas on the tower shown in Figure 4 serve both to communicate with the surface array stations and to relay surface and fluorescence detector data to the Auger Campus.

Telescope calibration and atmospheric monitoring are essential for precision energy measurement by fluorescence detectors. An end-to-end calibration uses a uniformly illuminated drum that covers the telescope aperture to calibrate the phototubes. The illumination of the drum is compared to a standard photodiode from the US National Institute for Standards and Technology. Instruments for atmospheric monitoring include lasers, lidar systems, horizontal attenuation monitors, cloud monitors, star monitors and balloon borne radiosondes. One of the most useful instrument systems for atmospheric monitoring, as well as other performance checks is the Central Laser Facility (CLF) [6]. For monitoring the atmosphere the CLF is equipped with a steerable laser that can be seen by the three operating fluorescence buildings. A particularly useful feature is achieved by connecting an optical fiber from the laser to a nearby surface detector station. The laser and the triggered station together can simulate hybrid events for timing and performance checks.

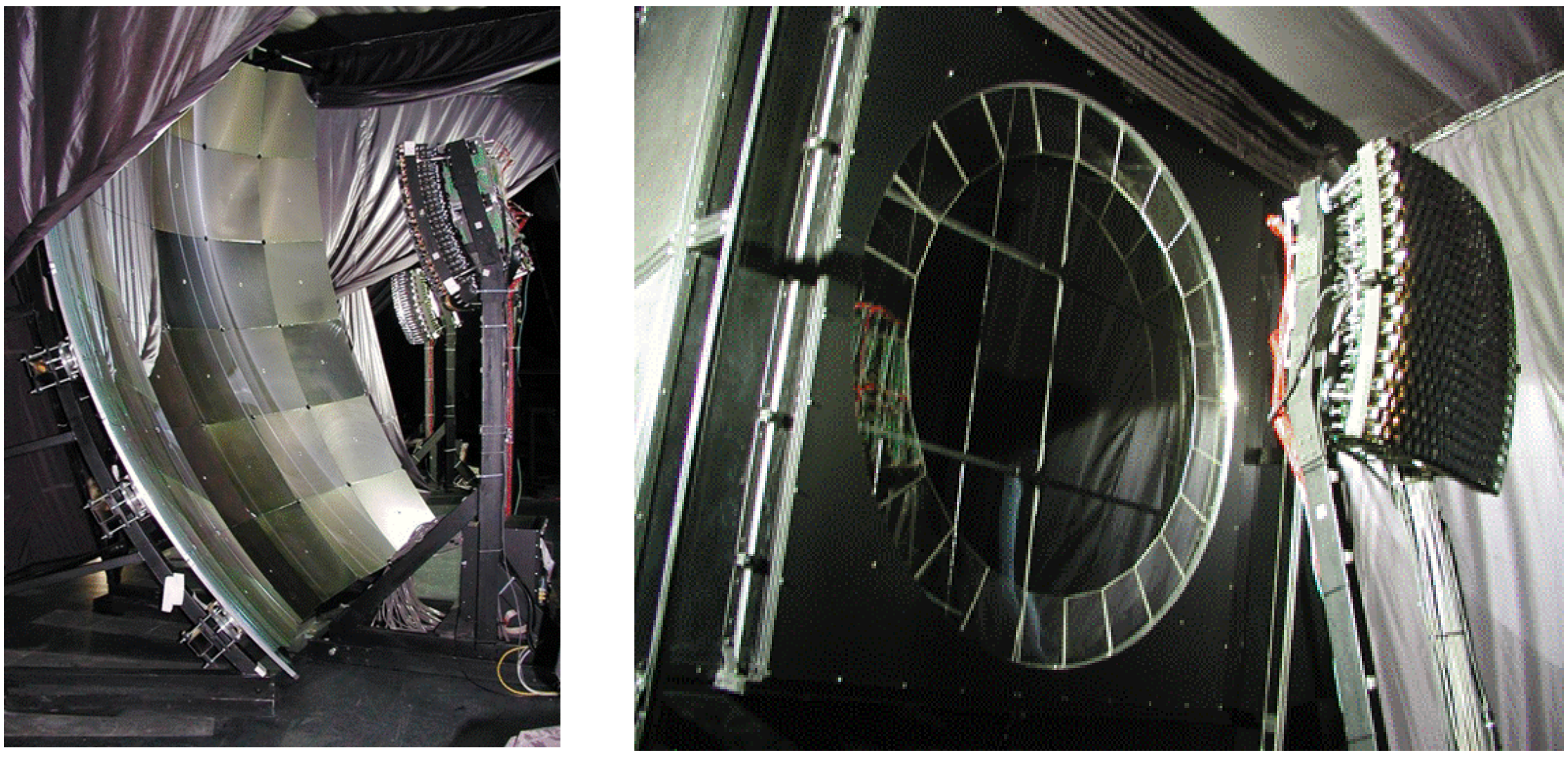

**Figure 3.** Two views of the fluorescence telescope. On the left is the 3.4 m diameter segmented mirror and the right is a view of the aperture stop with corrector lens and filter window. Each view shows the 440 phototube camera.

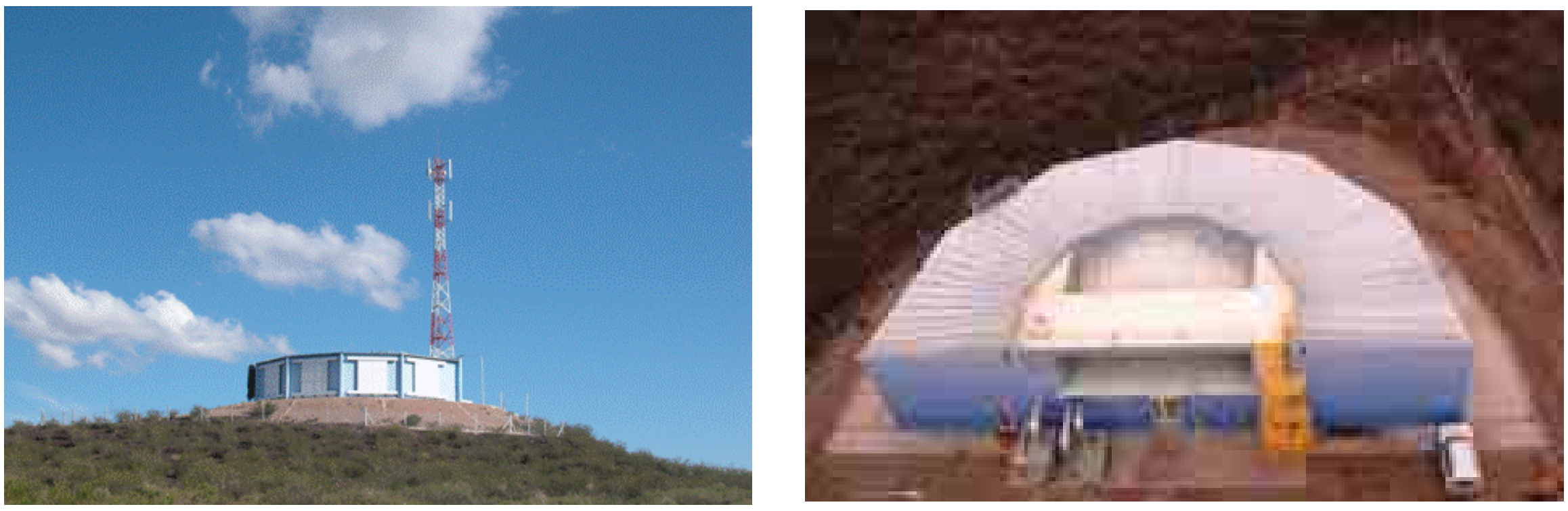

**Figure 4.** Two views of one of the fluorescence telescope enclosures. The view on the left includes the enclosure with shutters and the communications tower in the background. The view on the right is taken from the tower and shows the 6 – 30º telescope bays.

## 4.0 Construction progress

As of August 2005 there were 905 of the 1600 surface detector stations deployed. Of these, 835 were equipped with electronics and sending triggers to the central data acquisition system. Three of the four fluorescence detector buildings each with six telescopes are fully operational. The construction of the last building will begin within a few months. The communications tower has been erected at this site and is actively linking surface detectors in the area with the data acquisition system at the Auger campus. Observatory completion is expected in 2006.

## 5.0 Performance

Two air shower events will serve to illustrate the performance of the detectors [7, 8, 9]. The first of these is shown in Figure 5. It is an event of fairly high energy – about 70 EeV – at a moderate zenith angle of 48°. On the left is a typical flash ADC trace from one of the 18 detector stations participating in the event. The detector is about 1 km from the core. In the center panel is the pattern of hits on the array with the size of the spot is proportional to the log of the signal in that station. The arrow shows the reconstructed shower direction and core position. The last frame shows the lateral distribution of detector signals from the core position.

The next example is that of a hybrid event with energy of about 8 x $10^{18}$ eV shown in Figure 6. In the panel on the upper left is the array hit distribution. The solid line through the array represents the intersection of the plane formed by the shower and the fluorescence detector with the array. The core position as determined by surface array reconstruction is represented by the tip of the arrow. Note that the core position from the surface array lies close to the line formed by the shower/detector plane. The lateral density distribution in the surface array for this event is shown in panel on the upper right. The light collected by the fluorescence detector converted to energy deposition is shown in the lower left. The maximum of the shower development is clearly visible. The geometrical reconstruction of this event used both fluorescence detector and surface array information. This is shown by the plot in the panel on the lower right. The timing from the fluorescence pixels is plotted as the shower sweeps out the angle $\chi$ formed by the developing shower and the fluorescence detector. The times recorded as the shower front passes the surface array stations project directly on the fluorescence pixel curve.

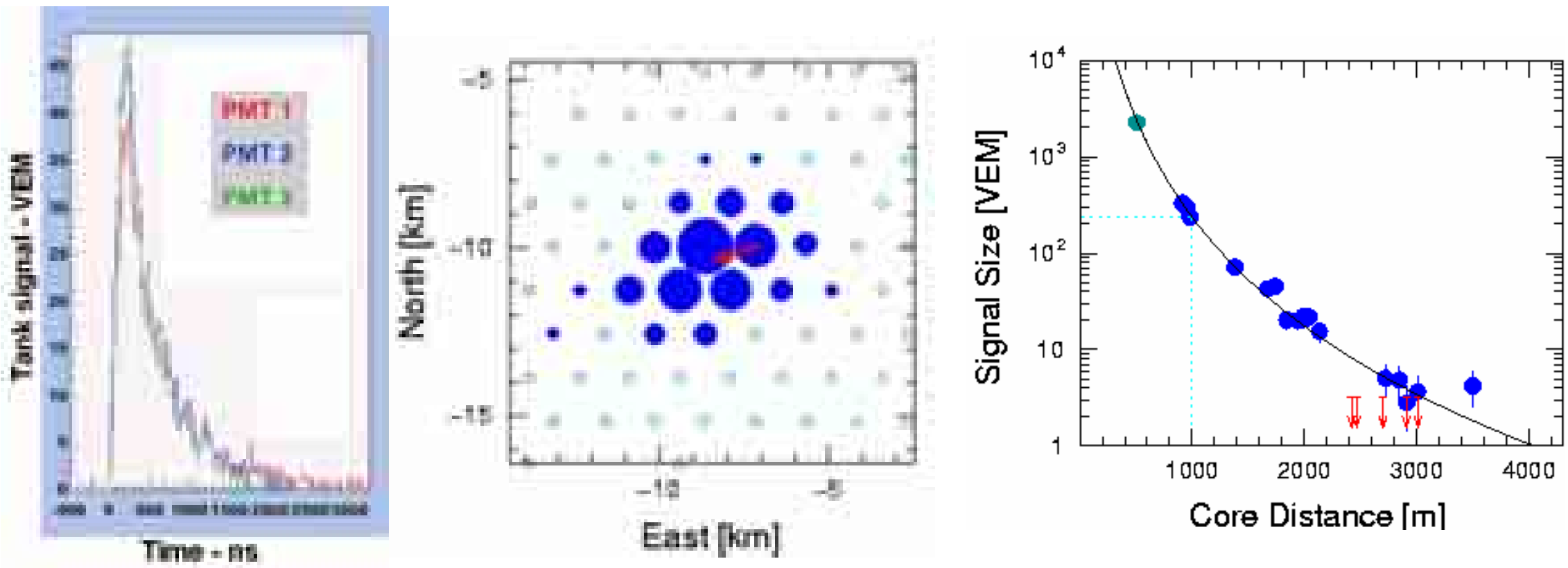

**Figure 5.** An example of a surface array event with an energy of 70 EeV and zenith angle of 48°. On the left is a typical FADC trace from one of the triggered stations located 1 km from the core. In the center is the hit pattern on the array with the axes in kilometers. On the right is the lateral density distribution of the event. The arrows correspond to stations that did not contribute to the trigger. The dotted line indicates the signal size at 1000 m or S(1000).

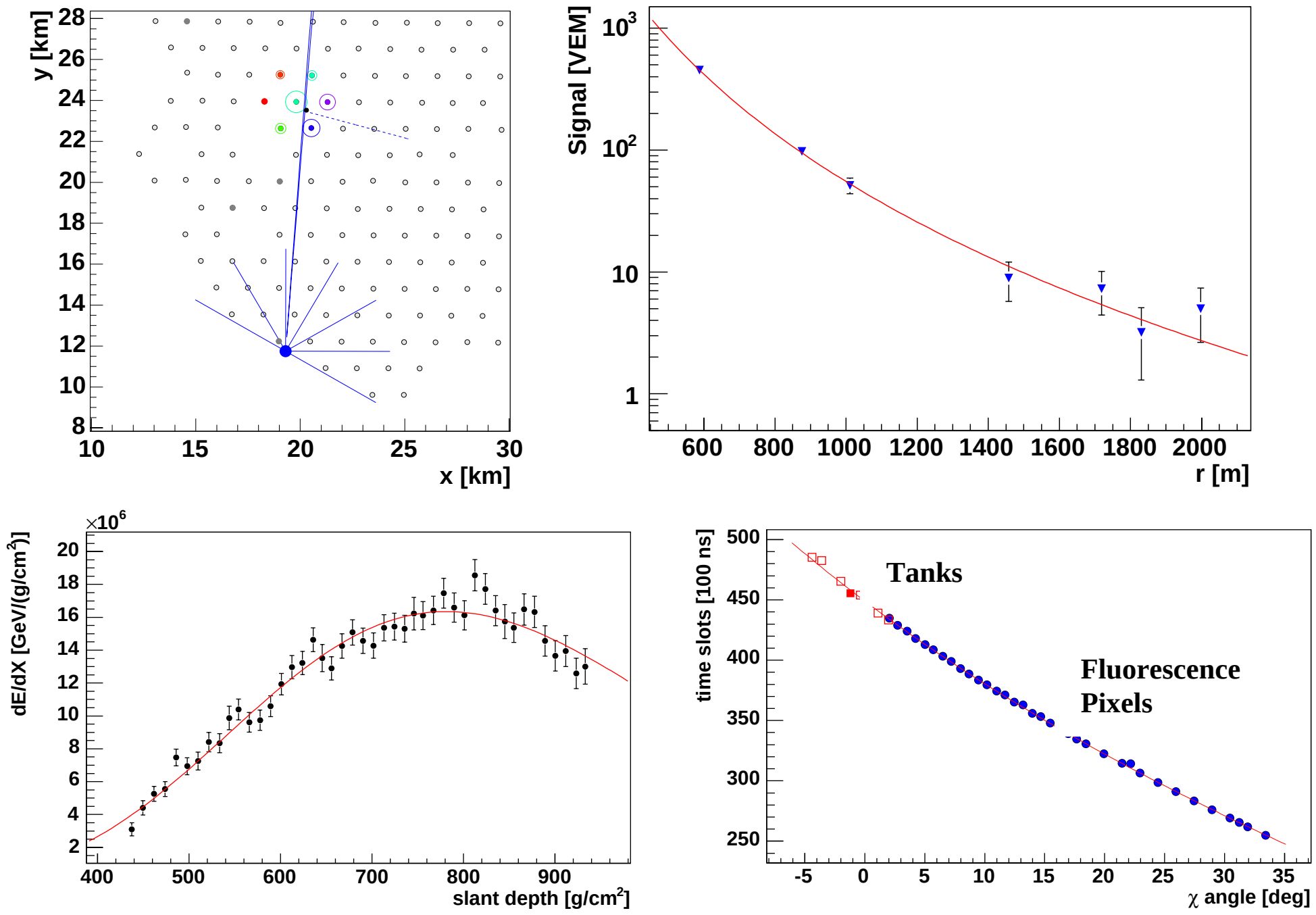

**Figure 6.** An example of a hybrid event. In the panel on the upper left is the array hit distribution. The solid line through the hit tanks represents the intersection of the plane formed by the shower and the fluorescence detector with the array. The lateral density distribution in the surface array for this event is shown in panel on the upper right. The panel on the lower left shows the energy deposition in the developing shower as measured by the fluorescence detector. In the panel in the lower right, fluorescence pixel hit times are shown as the shower sweeps out the angle $\chi$ formed by the developing shower and the fluorescence detector.

Another strength of the hybrid technique is the ability to understand the detailed performance of the detectors, for example the angular resolution and the core position resolution of the surface array [10]. The central laser facility is used to produce simulated hybrid events using the CLF laser and a surface array station nearby triggered by light from the same laser. These events are analyzed by the normal hybrid reconstruction to determine hybrid angular and core position resolution. This procedure yields a hybrid angular resolution 0.6 º (mean) within a 68% confidence level. The hybrid advantage is dramatically shown in Figure 7. The angle of the laser beam measured with the hybrid technique is the sharply peaked histogram and, in the background, is the reconstructed angle determined by the fluorescence detector alone. Once the hybrid angular resolution is known, the surface array only angular resolution can be extracted using the set of real hybrid events by analyzing the difference between the angle determined by hybrid reconstruction and the angle measured by surface array only. By this method the surface detector angular resolution was found to be better than 2.2º for 3-fold events (E < 4 EeV), better than 1.7 º for 4-folds events (3 < E < 10 EeV) and better than 1.4 º for higher multiplicity (E > 8 EeV) [11].

In a similar way the hybrid core position resolution can be determined by the simulated hybrid events provided by the CLF laser beam (Figure 7, right panel). The laser beam position determined by the hybrid reconstruction is shown in Figure 8 together with the laser position using the fluorescence detector only. The resulting hybrid core position resolution is found to be 57 m while the core position resolution for monocular reconstruction is about 566 m. As in the case of the angular resolution, the surface detector only core position resolution can be extracted from the hybrid data set and is about 150 m.

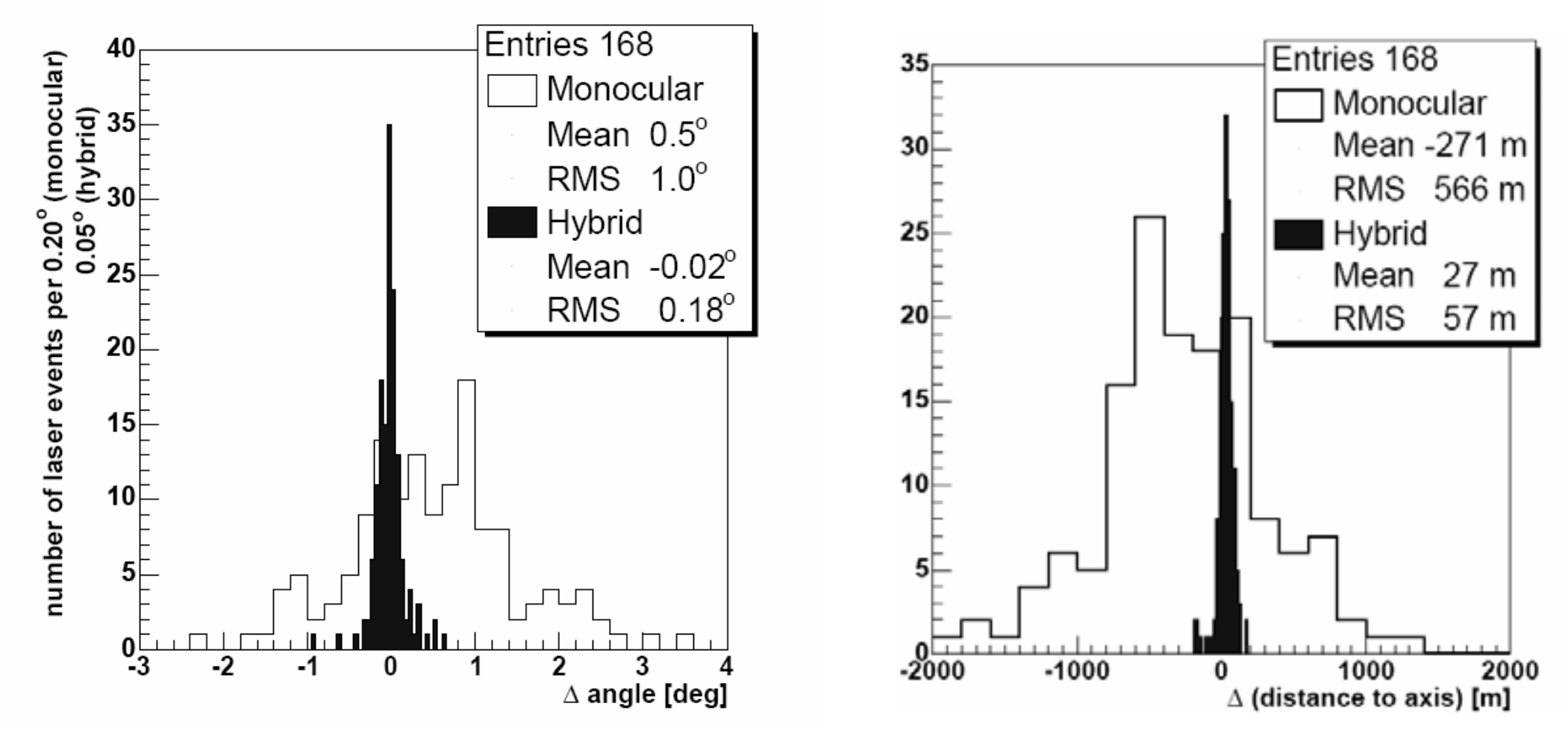

**Figure 7.** Determination of the hybrid angular and core position resolutions using laser simulated hybrid events. The space angle of the laser beam determined by the hybrid reconstruction is shown in the left panel. The broader plot in the background is the angle determined by the fluorescence detector only. The position of the laser "core" determined by hybrid reconstruction and by the fluorescence detector only are shown in the panel on the right. Both plots are illustrative of the power of fluorescence detector – surface array hybrid technique.

## 6.0 Energy Scale and the Spectrum

The first data set was collected from 1 January 2004 through 5 June 2005. Zenith angles of the events were restricted to lie between 0º and 60°. The acceptance was about 1750 km$^2$ sr yr. After suitable quality cuts, ~ 180,000 surface array events and ~ 18,000 hybrid events were recorded. For the first analysis of this data we have chosen to base the energy scale on fluorescence measurements in the hybrid event set. Aside from a 10% correction for the energy of muons and neutrinos lost in the ground, the energy scale and resulting spectrum do not depend on specific interaction models or assumptions about primary composition.

Energy determination by the surface array by itself relies on the measured particle density at a specified distance from the shower core. This particle density, given in terms of the signal in the ground array in units of VEM, is called the ground parameter S(r). We have chosen the signal size at 1000 m from the core – S(1000) – as a measure of the shower size which can be related to the primary energy.

A set of hybrid events was selected in which the fluorescence track length was at least 350 g $cm^{-2}$ and Cherenkov contamination was less than 10%. For each of these events the ground parameter S(1000) was compared with the fluorescence energy determined for that event. The ground parameter S(1000) was obtained for each event by non-linear interpolation along the LDF (see Figure 5). S(1000) was then corrected for zenith angle using the constant intensity method (see reference [12] for details). The resulting plot of log S(1000) vs. the log of the fluorescence determined energy for that set of hybrid events is shown in Figure 8. The energy converter used for assigning energies to surface array events is determined from the slope of a linear fit. Our rapidly increasing data set and expected reduction in systematic uncertainties will dramatically improve the precision of the energy scale.

The continuously operating surface array provides an unambiguous exposure. Calculation of the precise acceptance, however, has to be done with some care, given the dynamic size and shape of the array during construction [13]. The approach, however, has been quite conservative. Only shower events that are well away from the array boundaries and that have a full complement of active detectors near the core were used. The surface detector energy distribution of events divided by the acceptance yields the spectrum shown in Figure 9. The errors on these data points are statistical only. Since our hybrid based energy converter can now only be determined by our limited sample of hybrid events, the systematic uncertainty used in this method grows substantially at 100 EeV. Systematic uncertainties are estimated in two energy regions - $10^{18.5}$ eV and $10^{20}$eV. The horizontal bar shows the systematic energy uncertainty and the vertical bar is the uncertainty in the acceptance determination (10%). Because the hybrid events are used to define the energy converter, systematic uncertainties are dominated by the uncertainties in the fluorescence yield (15%) and the absolute detector calibration (~12%). The total FD systematic uncertainty from all sources is 26%.

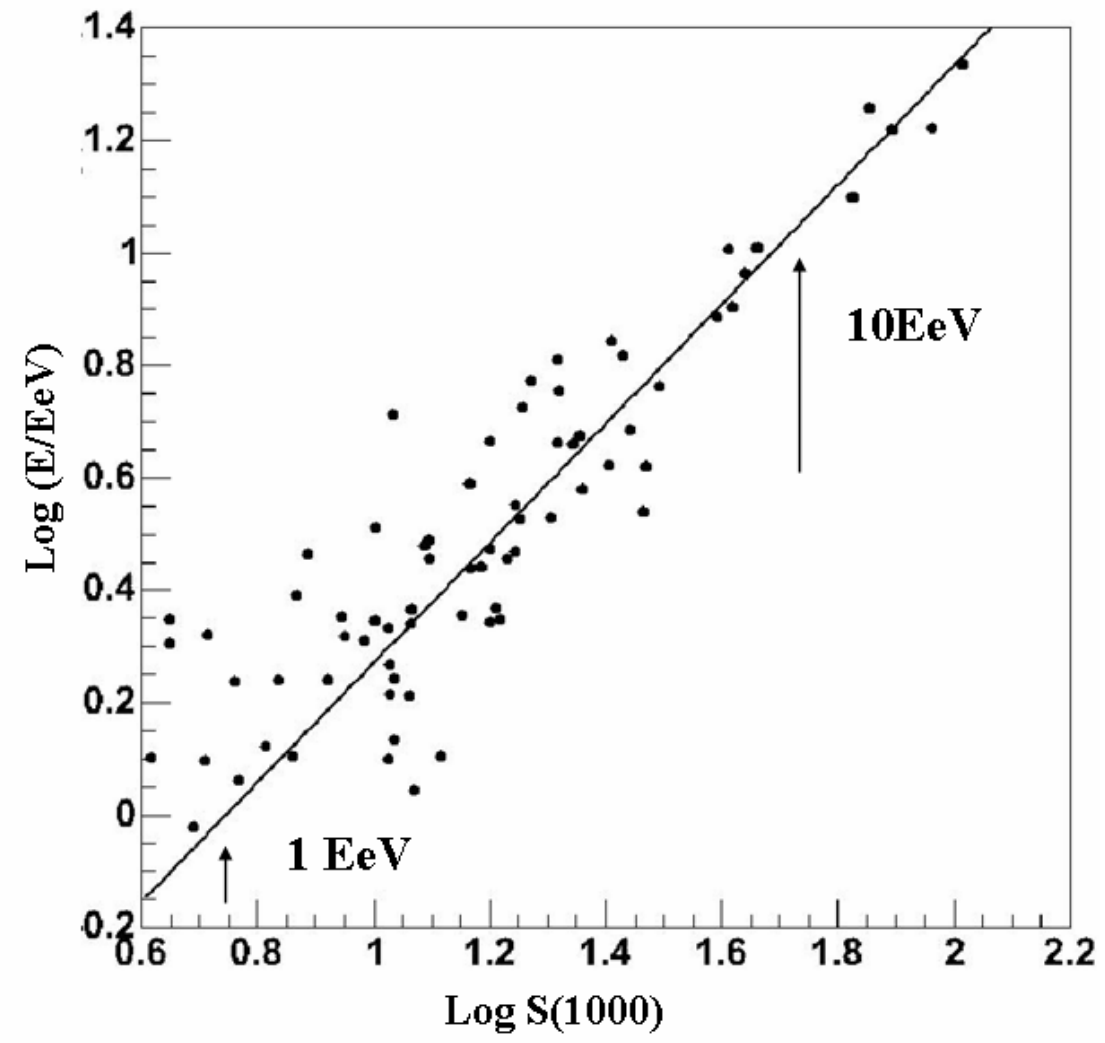

**Figure 8.** The plot of log S(1000) vs. the log of the fluorescence determined energy for a set of hybrid events.

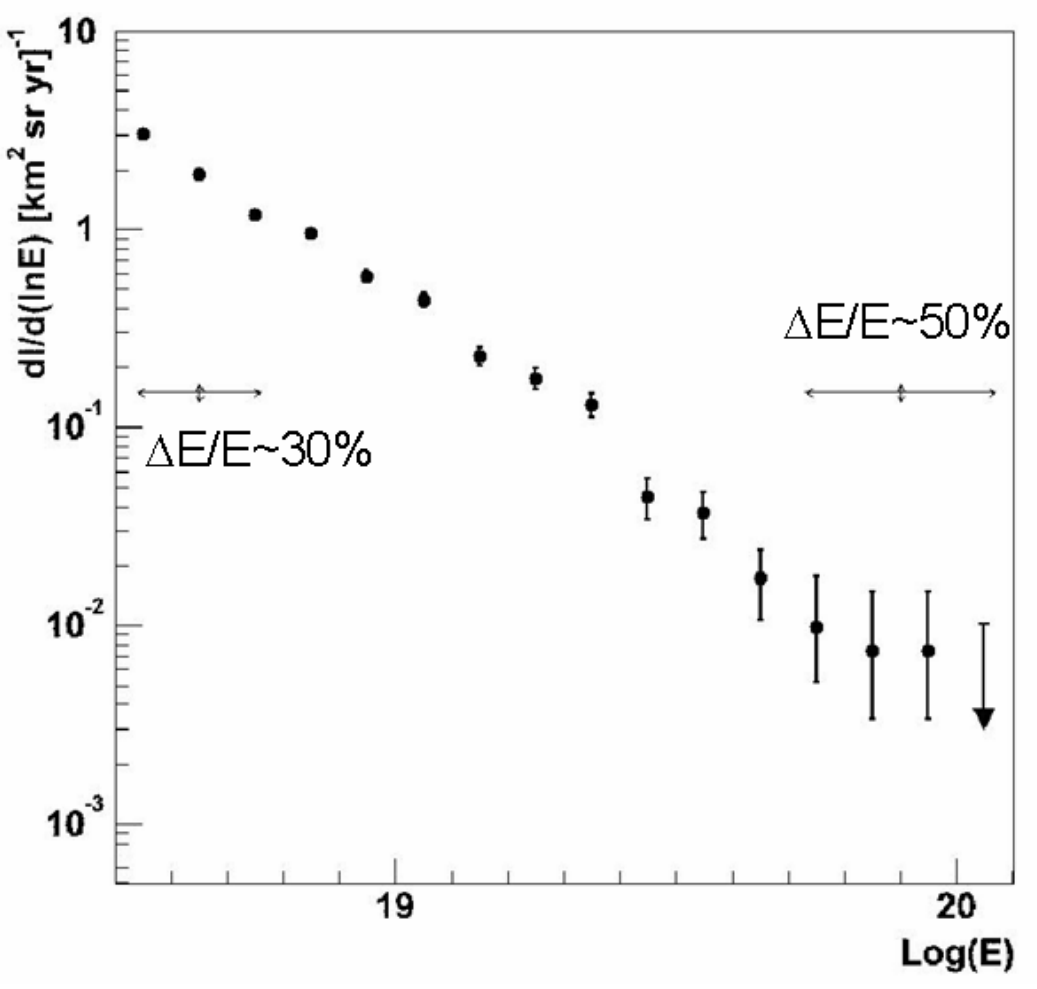

**Figure 9.** The Auger energy spectrum

Our preliminary results indicate that current statistical and systematic errors cannot distinguish between a continuing spectrum and one with GZK suppression. The number of events above $10^{20}$ eV is often used naively to indicate consistency (or not) with evidence of a GZK suppression. Given the uncertainty in energy, at least in the case of our results, a significant number of events about $10^{20}$ eV may still be consistent with a GZK feature while no events above $10^{20}$ eV may still be consistent with a continuing spectrum. Note that no events above $10^{20}$ eV are included in the spectrum. We did record a hybrid event with energy of 140 EeV (assuming only a pure Rayleigh atmosphere) but unfortunately the event was just outside that array and did not meet our criteria for inclusion in the spectrum (see reference [9] for a brief description of this event).

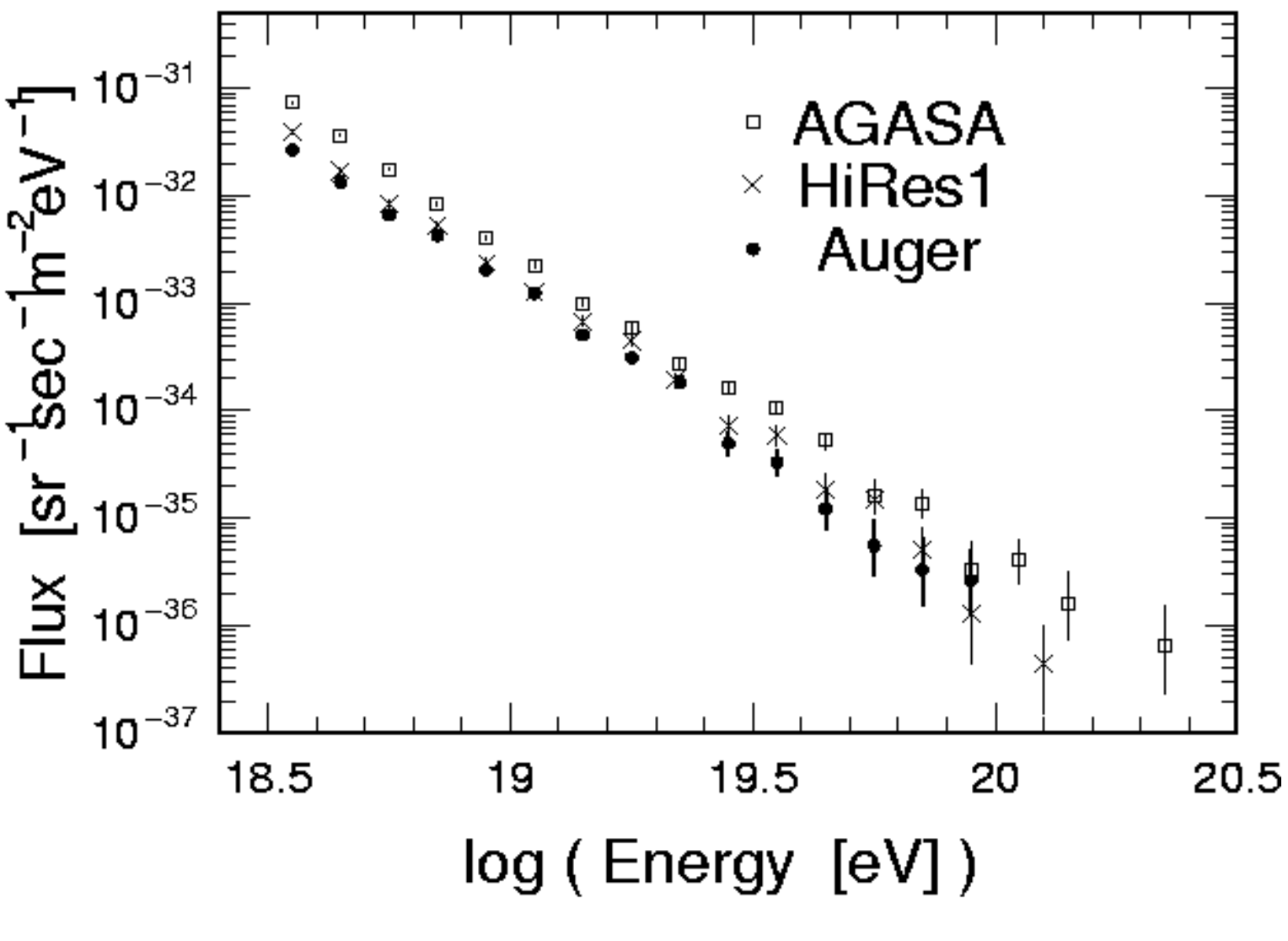

**Figure 10.** Comparison of the Auger spectrum with those of AGASA and HiRes 1 (mono).

Figure 10 shows the comparison of the Auger spectrum with that of AGASA [14] and HiRes 1(mono) [15]. The integrated aperture of Auger is near that of AGASA but less than that of HiRes1.

## 7.0 Summary

The Auger Observatory is now well over half finished. With only 25% of a full Auger-year exposure we have defined an analysis strategy and have produced a model-independent spectrum. We have also performed our first studies of anisotropies in the southern sky [16,17] and have set limits on photon primaries[18].

Over the next few months we will use the rapidly expanding data set to improve the energy assignment, reduce systematic uncertainties and endeavor to more fully understand our instruments. In two years' time we will increase our data sample by a factor of five to seven. This will enable a high statistics study of the spectrum in the GZK region, anisotropy and point source searches and composition studies. We also intend to exploit events with zenith angles above 60 degrees. Quasi-horizontal air showers will be of particular interest as such events are likely to be initiated by ultrahigh energy cosmic neutrinos.